\documentclass[10pt,final,twocolumn,aps,pra,floatfix]{revtex4-1}
\usepackage[utf8]{inputenc}
\usepackage{graphicx}
\usepackage{color}
\usepackage{subscript}
\usepackage{amsmath,amssymb}

\begin{document}

\title{Kinetic analysis of negative power deposition in low pressure plasmas}

\author{Jan Trieschmann}
\author{Thomas Mussenbrock}

\affiliation{Ruhr University Bochum, Department of Electrical Engineering and Information Science, 44780 Bochum, Germany}

\date{\today}

\begin{abstract}
The negative power absorption in low pressure plasmas is investigated by means of an analyical model which couples Boltzmann's equation and the quasi-stationary Maxwell's equation. Exploiting standard Hilbert space methods an explicit solution for both, the electric field and the distribution function of the electrons for a bounded discharge configuration subject to an unsymmetrical excitation has been found for the first time. The model is applied to a low pressure inductively coupled plasma discharge. In this context particularly the anomalous skin effect and the effect of phase mixing is discussed. The analytical solution is compared with results from electromagnetic full wave particle in cell simulations. Excellent agreement between the analytical and the numerical results is found.
\end{abstract}

\maketitle

\section{Introduction}
Low temperature plasmas have found widespread use in materials processing. Particularly inductively coupled plasmas (ICPs), which are in the spotlight of this paper, play a crucial role for a number of applications, e.g., for anisotropic etching of sub-micron patterns.\cite{keller1993, hopwood1993, ventzek1994, panagopoulos2002, lieberman2005} Surface modifications like anisotropic etching require that the impinging ions do not significantly experience collisions with the neutrals of the background gas, in particular in the region above the substrate or wafer. This can only be achieved in the low gas pressure regime, where the gas pressure is not higher than 10 mTorr. In this so called non-local regime the local relationship between current density and electric field in the plasma described by a simple Ohm's law given by 
\begin{gather}\label{localohm}
\vec j(\vec r,t)=\sigma(\vec r, t) \vec E(\vec r, t)
\end{gather}
is not valid anymore. The reason is that it assumes the thermal motion of electrons to be negligible and that the plasma is heated by local collisional dissipation of electromagnetic energy. The collisions of electrons with the background gas perturb their regular motion which leads to frequent randomization, effectively erasing the electrons' memory. The phase is reset on every collision encounter and there is no phase shift between the current density and the electric field in between two encounters. The heating therefore results in local Ohmic power dissipation.

The situation at low gas pressure is completely different. Since the electron mean free path is comparable or even larger than the size of the plasma itself, collisions of electrons with the background gas become rare. Consequently, local collisional heating ceases to be an effective mechanism. The power dissipation is mainly due to collisionless electron heating.\cite{turner1993, godyak1998, cunge2001,lafleur2015} A phase shift establishes due to the finite thermal velocity of the electrons. That is, thermal electrons move through the plasma and gain electromagnetic energy from the electric field particularly in the confined region close to the boundary, i.e., the so-called skin layer. If the thermal velocity of the electrons is high enough as to pass through the skin layer on a time scale short compared to the time scale of the oscillating electromagnetic field, then the moving electrons may gain net energy from the electromagnetic field. The required randomization of the regular electron motion is again due to collisions, however, the heating is non-local.

In a non-local picture, one can find that the electrical field $\vec E$ at the position $\vec r^{\,\prime}$ inside the plasma at an instant of time in the past $t^\prime$ influences the current density $\vec j$ at the position $\vec r$ at present time $t$. The local version of Ohm's law has to be rewritten,
\begin{align}
\label{ohmnonlocal}
\vec j(\vec r,t)=\int_V \int_{-\infty}^t \sigma(\vec r, \vec r^{\,\prime}, t-t^\prime) \vec E(\vec r^{\,\prime}, t^\prime) d^3r^\prime dt^\prime.
\end{align}
Here, $\sigma$ is now a distributed conductivity containing all the necessary information from all past instants of time at all positions in space.

It has been long acknowledged that the spatial and temporal dispersion of the plasma conductivity is due to a ``warm plasma'' (i.e., the thermal motion of electrons), analogous to the anomalous skin effect in metals.\cite{pippard1949, mattis1958} In a plasma the anomalous skin effect is first discussed by Weibel.\cite{weibel1967} Certain peculiarities are observed in both experiments and theory: The distribution of the electromagnetic field shows a non-monotonic decrease, rather than an exponential decrease of the classical skin effect, so that the appearance of local minima and maxima of the electromagnetic field are observed. The most peculiar phenomenon reported is however negative power deposition. This means that the electrons gain kinetic energy from the electric field at some position and time and later return a fraction of their kinetic energy to the electrical field at a different position.\cite{reynolds1969, kofoid1969, blevin1973, yoon1996, godyak1997, kolobov1997}

The anomalous skin effect in gaseous plasmas is particularly studied using one-dimensional models for both, bounded and infinite domains. Most results reported for bounded plasmas assume a uniform plasma density. Kaganovich et al.\ developed self-consistent one-dimensional models in order to study the power deposition and electron heating in ICPs.\cite{ramamurthi2003a, kaganovich2004, kaganovich2006, polomarov2006} Particularly, the effect of the electron energy distribution function on the power deposition and the plasma density is studied by means of a self-consistent one-dimensional model.\cite{ramamurthi2003b, froese2009} Quite recently, Hagelaar reports on a fluid description of non-local electron kinetics in inductively coupled plasmas in order to gain scaling laws for the non-monotonic spatial structure of the electromagnetic field, the anomalous skin effect, and a condition for the appearance of local negative power deposition.\cite{hagelaar2008a, hagelaar2008b} 

To the best of our knowledge, non of the approaches provide a self-consistent analytic solution for a one-dimensional bounded domain, unsymmetrical with respect to the boundary conditions (i.e., with excitation only from one side). In this work we focus exactly on this problem. We analyze the anomalous skin effect and negative power deposition in an unsymmetrical plasma at low pressure using an analytical kinetic model of a finite plasma slab. The paper is organized as follows: We first revisit the basic equations derived from consistently coupled Maxwell's equations and Boltzmann's equation, and motivate the spatially one-dimensional description of the problem. We formulate a boundary value problem of Sturm-Liouville type for the electron distribution function and the penetrating electric field. Boltzmann's equation is solved using an analytical ansatz similar to the approach proposed by Tyshetskiy et al.\cite{tyshetskiy2002} However, we do not assume an exponentially decreasing electrical field in an infinite plasma, but we develop self-consistent analytical solutions to both problems -- the distribution function and the electrical field -- by means of Hilbert space methods. We finally discuss the results for certain generic plasma parameters (for a low-pressure case) and compare them with results from particle in cell simulations. Particularly, we discuss phase mixing, negative power deposition, and the effect of a finite thermal electron velocity on the distribution of the electrical field and the power deposition.

\section{Basic equations}

In order to describe the anomalous skin effect, which is a spatially and temporally non-local effect, Maxwell's equations have to be coupled consistently to Boltzmann's equation for each constituent of the plasma. Of course, this set of equations is not tractable. It has to be simplified by means of a scale analysis.

We assume the angular (radio-)frequency of the penetrating electromagnetic field $\omega$ to lie significantly above the ion plasma frequency $\omega_{pi}$ and significantly below the electron plasma frequency $\omega_{pe}$. For the time scales we assume the ordering $\omega_{pi}\ll\omega\ll\omega_{pe}$. The ions are not affected by the rf modulation and react only on the phase-averaged electric field. The radio-frequency current is therefore carried by electrons alone. To describe the electrodynamic effects it is therefore justified to formulate Boltzmann's equation solely for the electrons
\begin{gather}
\label{bg11}\frac{\partial f}{\partial t}+\vec v\cdot\nabla f - \frac{e}{m}\left(\vec E +\vec v\times \vec B\right)\cdot \nabla_{\! v} f=\left.\frac{\partial f}{\partial t}\right|_{  col},
\end{gather}
where $f$ is the distribution function of electrons and $m$ and $-e$ are the electron mass and charge, respectively. The plasma is assumed to be quasi-neutral and (for simplicity) homogeneous, $n_{e}=n_{i}=n$. The plasma density is given by
\begin{gather}
\label{bg12}n=\int f d^3 v.
\end{gather}
Maxwell's equations governing the electric field $\vec E$ and the magnetic field $\vec B$ can then be written in the quasi-stationary approximation, 
\begin{align}
\label{mg11}\nabla\times\vec B &= \mu_0 \vec j,\\
\label{mg12}\nabla\times\vec E &= -\frac{\partial\vec B}{\partial t},\\
\label{mg13}\nabla\cdot \vec B &=0,\\
\label{mg14}\nabla\cdot \vec E &=0.
\end{align}
Here, the displacement current is neglected.\cite{mussenbrock2008} This approximation is justified since it scales with $\omega/\omega_{pe}$. The current density is defined as the first velocity moment of the distribution function
\begin{align}
\label{currentdensity1}\vec j = -e \int \vec v f d^3 v.
\end{align}

This system of coupled nonlinear partial differential equations is still very difficult to solve. We therefore study the perturbation of the equilibrium state and apply a linearization of the dynamical (and time-harmonic) quantities (indicated by tilde). We set $f=\bar f+\tilde f$ with the Maxwellian distribution $\bar f$, the magnetic field $\vec B=\bar{\!\vec B}+\tilde{\!\vec B}=\tilde{\!\vec B}$, 
and the electrical field  $\vec E=\bar{\!\vec E}+\tilde{\!\vec E}=\tilde{\!\vec E}$. 
Since we are interested in the high-frequency dynamics of the system, we set the equilibrium electromagnetic field given by $\bar{\!\vec B}$ and $\bar{\!\vec E}$ to zero. We then obtain a system of linear equations for the dynamical quantities,
\begin{align}
\vec v\cdot 
\nabla \tilde f  + \left( \nu + i \omega \right)\tilde f &=
 \frac{e}{m}\tilde{\!\vec E} \cdot\nabla_{\!v}\, \bar f,\\
\nabla\times\nabla\times\tilde{\!\vec E}&= i\omega e \mu_0 \int \vec v\, \tilde f d^3 v.
\end{align}
This system of differential equations has to be solved subject to appropriate boundary conditions for both, the distribution function $\tilde f$ and the electromagnetic field. In the given system the collisions of electrons with the neutrals of the background gas are taken into account using an effective collision rate for momentum transfer described by $\nu_{m}$. The divergence equations \eqref{mg13} and \eqref{mg14} act as constraints. They are given by
\begin{align}
\nabla\cdot\tilde{\!\vec B}&=0,\\
\nabla\cdot\tilde{\!\vec E}&=0.
\end{align}
Additionally, we have
\begin{gather}
\int \tilde f d^3 v  = 0.
\end{gather}

\begin{figure}[t]
\includegraphics[width=7cm]{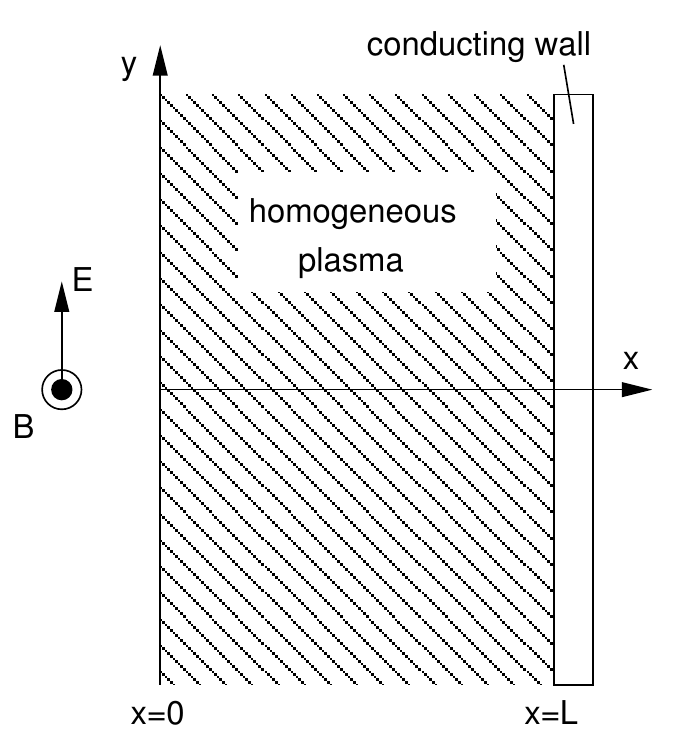}
\caption{One-dimensional plasma slab geometry.}
\label{fig01}
\end{figure}

In order to analytically study the anomalous skin effect and related dissipation mechanisms we investigate a one-dimensional plasma slab as depicted in figure~\ref{fig01}. An electromagnetic field given by $\tilde{\!\vec B}=\tilde{\! B}(x)\vec e_{z}$ and $\tilde{\!\vec E} =\tilde{\! E}(x)\vec e_{y}$ prescribed at $x=0$ penetrates into a semi-infinite bounded homogeneous low-pressure plasma of the length $L$. It is important to notice that the plasma is infinite in the lateral plane, but bounded in $x$ direction. For this one-dimensional case the differential equations become
\begin{align}
\label{linboltz}
\left(\frac{\partial}{\partial x}  + \frac{\alpha}{v_{x}} \right)\tilde f&= \frac{\beta}{v_{x}}  \tilde E(x),\\
\label{blahelm}
\frac{\partial^2 \tilde E}{\partial x^2} &= i\omega  \mu_0 \,\tilde{ \! j}(x;\tilde{\! E})
\end{align} 
with $\alpha=i\omega+\nu_{m}$ and $\beta=-e \bar f v_{y}/ m v^2_{th}$. The thermal velocity of the electrons is assumed to be $v_{th}=(T_{e}/m)^{1/2}$ with the electron temperature $T_{e}$ (in eV; the electron mean velocity is $\bar{v}=(8 T_{e}/\pi m)^{1/2}$). The high-frequency current density $\vec \tilde{\! j}=\tilde{\! j}(x) \vec e_{y}$, which is implicitly a function of the electrical field (and the position) is given by the first moment of the dynamical distribution function $\tilde f$,
\begin{align}
\tilde{\! j}=- e \int v_{y} \tilde f(x,\vec v;\tilde{\! E}) d^3 v.
\end{align}
As boundary conditions for \eqref{blahelm}, we assume a time-harmonic magnetic field $B_0$ oscillating with angular frequency $\omega$ given at $x=0$ and let the electric field vanish at the conducting boundary at $x=L$. The boundary conditions for the electric field $\tilde E$ are therefore given by
\begin{align}
\label{boundaryconditions1} \left.\frac{\partial\tilde{\! E} }{\partial x}\right|_{x=0}&= - i\omega B_0,\\
\label{boundaryconditions2} \left. \tilde{\! E}\right|_{x=L} &= 0.
\end{align}
Gathering equations \eqref{blahelm} to \eqref{boundaryconditions2}, we have a well-posed boundary value problem for the electric field, i.e., a homogeneous differential equation and inhomogeneous boundary conditions.

For practical reasons it is convenient to have homogeneous boundary conditions for the electric field. We thus apply for the electric field the ansatz
\begin{align}
\label{ansa}
\tilde{\! E} = E_0 + E_{h},
\end{align}
with $E_0=i\omega (L-x)B_0$. The boundary value problem for $E_{h}$ becomes
\begin{align}
\label{bla}
\frac{\partial^2 E_{h}}{\partial x^2} = i\omega  \mu_0 \,\tilde{ \! j}(x;  E_0 + E_{h}),
\end{align}
subject to homogeneous boundary conditions
\begin{align}
\label{rwbc1}
\left.\frac{\partial E_{h}}{\partial x}\right|_{x=0} &= 0\\
\label{rwbc2}
\left. E_{h} \right|_{x=L} &= 0.
\end{align}
Since \eqref{linboltz} is linear, the right hand side of \eqref{bla} can be modified. We can decompose the current density in two components, $\,\tilde {\! j}(x;  E_0 + E_{h})=j_0(x; E_0) +j_{h}(x; E_{h})$, where $j_0$ is due to $E_0$ and $j_{h}$ is due to $E_{h}$. The two current densities are then given by
\begin{align}
\label{jot0}
j_0&= -e \int v_y f_0(x,\vec v;E_0) d^3 v,\\
\label{jot}
j_{h}&= - e \int v_y f_{h}(x,\vec v;E_{h}) d^3 v,
\end{align}
with the corresponding decomposition from $\tilde f = f_0 + f_{h}$. 

The distribution functions $f_0$ and $f_{h}$, which implicitly depend on the electric field $E_0$ and $E_{h}$, are the solutions to the linearized Boltzmann equation \eqref{linboltz} given the appropriate right hand side inhomogeneity, which is
\begin{align}
\label{eqf0}
\left(\frac{\partial}{\partial x}  + \frac{\alpha}{v_{x}} \right) f_0(x,\vec v;E_0)&= \frac{\beta}{v_{x}}  E_0(x),\\
\label{eqf}
\left(\frac{\partial}{\partial x}  + \frac{\alpha}{v_{x}} \right) f_{h}(x,\vec v;E_{h})&= \frac{\beta}{v_{x}} E_{h}(x).
\end{align}
The (spatial) boundary conditions for the distribution functions are so-called ``reflecting'' boundary conditions with
\begin{align}
\label{rbv0}
\left. f_0(v_{x})\right|_{x=0,L}&=\left. f_0(-v_{x})\right|_{x=0,L},\\
\label{rbv}
\left. f_{h}(v_{x})\right|_{x=0,L}&=\left. f_{h}(-v_{x})\right|_{x=0,L}.
\end{align}
We now have exactly \emph{one} first order differential equation for each distribution function $f_0$ and $f_{  h}$, each of which subject to exactly \emph{two} boundary conditions, one at either sides. This is in fact a well-posed mathematical problem which can be solved analytically, as shown the subsequent section.

With the knowledge on the distribution functions, and thus with the knowledge on the current densities, we are able to reformulate the desired inhomogeneous boundary value problem for the electrical field
\begin{align}
\label{rwp}
\frac{\partial^2 E_{h}}{\partial x^2} - i\omega  \mu_0 \, j_{h}(x; E_{h})= i\omega  \mu_0 \, j_0(x;  E_0).
\end{align}
subject to homogeneous boundary conditions \eqref{rwbc1} and \eqref{rwbc2}. This boundary value problem is of the Sturm-Liouville type and can also be solved analytically by means of standard Hilbert space methods.

\section{Analytical solution to the kinetic equations}

The first step toward the analytical solution of the electric field equation \eqref{rwp} is to calculate the current densities $j_0$ und $j_{h}$ from the solutions of the kinetic equations \eqref{eqf0} and \eqref{eqf}. These can be obtained by means of the method of integrating factors. Hence, the general solution to differential equations of the given type is
\begin{align}
f(x,\vec v; E) =& f(x_0)e^{-\alpha( x - x_0)/v_{x}} \nonumber\\ &+ \frac{\beta e^{-\alpha x/v_{x}}}{v_{x}} \int_{x_0}^x e^{\alpha x^\prime/v_{x}} E(x^\prime) dx^\prime.
\end{align} 
Here, we have \emph{one} unknown constant $f(x_0)$, but \emph{two} boundary conditions $\left. f \right|_{x=0,L}$.  This inherent problem can be overcome using the appropriate ansatz $f = f^+ +  f^-$ for the distribution function. $f^+$ represents the solution for $v_{x}\ge 0$, while vanishing for $v_{x}<0$, and satisfying the boundary condition at $x=0$. Accordingly, $f^-$ is the solution for $v_{x}\le 0$, while vanishing for $v_{x}>0$, and satisfying the boundary condition at $x=L$. We consequently obtain for $f^+$ and $f^-$, respectively
\begin{widetext}
\begin{align}
\label{generalsolutionf+}
f^+ &= \left\{\begin{array}{llll}\displaystyle
f(0)e^{-\alpha x/v_{  x} } + \frac{\beta e^{-\alpha x/v_{  x}}}{v_{  x}}\int_{0}^x e^{\alpha x^\prime/v_{  x}} E(x^\prime) dx^\prime\quad &\text{for}\quad v_{  x}\ge 0,\\
0\quad &\text{for}\quad v_{  x}< 0,
              \end{array}\right.\\
\label{generalsolutionf-}
f^- &= \left\{\begin{array}{llll}\displaystyle
f(L)e^{-\alpha( x - L)/v_{  x}} + \frac{\beta e^{-\alpha x/v_{  x}}}{v_{  x}}\int_{L}^x e^{\alpha x^\prime/v_{  x}} E(x^\prime) dx^\prime\quad&
\text{for}\quad v_{  x}\le 0,\\
0\quad&\text{for}\quad v_{  x}> 0.
                     \end{array}\right.
\end{align}
\end{widetext}
The unknown coefficients $f(0)$ and $f(L)$ can now be calculated explicitly from the boundary conditions for $f^+$ and $f^-$, which are
\begin{align}
\label{rbv+-}
\left.f^+(v_{  x})\right|_{x=0,L}=\left.f^-(-v_{  x})\right|_{x=0,L}.
\end{align}
We obtain the two expressions
\begin{widetext}
\begin{align}
f(0)&=\frac{\beta}{\left(1-e^{2\alpha L/v_{  x}}\right)v_{  x}}\left(e^{2\alpha L/v_{  x}} \int_L^0 e^{-\alpha x^\prime/v_{  x}} E(x^\prime) dx^\prime  - \int_0^L e^{\alpha x^\prime/v_{  x}} E(x^\prime) dx^\prime \right),\\
f(L)&=\frac{\beta e^{\alpha L/v_{  x}}}{\left(1-e^{2\alpha L/v_{  x}}\right)v_{  x}}\left( \int_L^0 e^{-\alpha x^\prime/v_{  x}} E(x^\prime) dx^\prime  - \int_0^L e^{\alpha x^\prime/v_{  x}} E(x^\prime) dx^\prime \right).
\end{align}
\end{widetext}
Herewith, a general analytic solution as a function of an arbitrary electric field $E(x)$ and subject to reflecting boundary conditions \eqref{rbv+-} is found. The solutions to \eqref{eqf0} and \eqref{eqf} have to be phrased in terms of $E_0$ and unknown $E_{  h}$, correspondingly. 

For the solution of the boundary value problem \eqref{rwp}, the current density $j_{  h}$ can now be calculated explicitly from \eqref{jot} (except for the integration over $v_{  x}$). It is convenient to again redefine the current density as
\begin{gather}
\label{jneu}
j_{  h}= - e \int_{-\infty}^\infty \mathcal{L}(x,v_{  x};  E_{  h}) d v_{  x}
\end{gather}
with
\begin{gather}
\label{operatorL}\mathcal{L}= \int_{-\infty}^\infty \int_{-\infty}^\infty  v_{  y} f_{  h}(x,\vec v;E_{  h}) d v_{  y}d v_{  z}.
\end{gather}
It is found that the integral over $v_{  x}$ converges and can thus be calculated by means of standard numerical methods \cite{eisenbarth2006}.

To obtain the right hand side of \eqref{rwp}, the current density $j_0$ has to be calculated as well. This can be done by replacing the electric field $E$ in the general solution \eqref{generalsolutionf+} - \eqref{generalsolutionf-} by the known expression $E_0=i\omega(L-x)B_0$. From \eqref{jot0} we consequently obtain
\begin{gather}
\label{j0neu}
j_0= - e \int_{-\infty}^\infty \mathcal{R}(x,v_{  x};  E_0) d v_{  x}
\end{gather}
with
\begin{gather}
\label{operatorR}\mathcal{R}= \int_{-\infty}^\infty \int_{-\infty}^\infty  v_y f_0(x,\vec v;E_0) d v_{  y}d v_{  z}.
\end{gather}
This expression can be written down explicitly. However, since it is quite cumbersome, we abstain from doing so \cite{eisenbarth2006}. Now all constituents of expression \eqref{rwp} are known and the boundary value problem for the electrical field $E_{  h}$ can be solved.

\section{Analytical solution to the electric field equation}

The boundary value problem for the electric field \eqref{rwp} and \eqref{rwbc1}-\eqref{rwbc2} is of Sturm-Liouville type. With the current densities given by \eqref{jneu} and \eqref{j0neu}, we obtain
\begin{align}
\label{rwpneu}
\frac{\partial^2 E_{  h}}{\partial x^2} + i\omega  \mu_0  e & \int_{-\infty}^\infty \mathcal{L}(x,v_{  x}; E_{  h}) d v_{  x}\nonumber \\ &= -i\omega  \mu_0  e \int_{-\infty}^\infty \mathcal{R}(x,v_{  x}; E_0) d v_{  x}.
\end{align}
with homogeneous boundary conditions $\partial_x E_{h}|_{x=0} = 0$ and $E_{h}|_{x=L} = 0$.
Such Sturm-Liouville problems are mathematically well characterized and can be solved by means of standard  Hilbert space methods: It is known that (i) the eigenfunctions of a Sturm-Liouville operator form a complete set of orthogonal functions, and (ii) the eigenvalues form a set of non-degenerated positive numbers. For the problem at hand, the assigned eigenvalue problem (differential equation subject to decoupled boundary conditions) reads explicitly
\begin{align}
\frac{\partial^2 u_{  k}(x)}{\partial x^2}&=\lambda_{  k}u_{  k}(x),\\[1ex]
\left.\frac{\partial u_{  k}}{\partial x}\right|_{x=0}&=\left.u_{  k}\right|_{x=L}=0.
\end{align}

The associated eigenfunctions $u_{  k}(x)$ and eigenvalues $\lambda_{  k}$ are defined for integer $k$ and given as follows
\begin{align}
u_{  k}(x)=\cos \sqrt{\lambda_{  k}} x,\\
\lambda_{  k}=\frac{(2k -1)^2 \pi^2}{4 L^2}.
\end{align}
Since the eigenfunction forms a complete set of orthogonal functions we can expand $E_{  h}$ into an infinite series
\begin{align}
\label{exp}
E_{  h}(x)= \sum_{k=1}^\infty a_{  k} \cos \sqrt{\lambda_{  k}} x.
\end{align}

In order to calculate the unknown coefficients $a_{  k}$ and thus the analytical solution of \eqref{rwpneu} it is convenient to expand the kernel of the integrals in terms of the same set of orthogonal functions. After plugging in the expansion of the electrical field \eqref{exp} into the kernel $\mathcal{L}$ given by \eqref{operatorL} we obtain
\begin{align}
\mathcal{L}(x,v_{  x})=&\sum_{k=1}^\infty a_{  k} \left(
b_{  k}(v_{  x})e^{-\frac{x \alpha}{v_{  x}}}\right. \nonumber\\ &+\left.
c_{  k}(v_{  x})\sin \sqrt{\lambda_{  k}} x +
d_{  k}(v_{  x})\cos \sqrt{\lambda_{  k}} x 
\right).
\end{align}
The coefficients $b_{  k}$, $c_{  k}$, $d_{  k}$ are themselves functions of $v_{  x}$. Finally, this expression for $\mathcal{L}$ has to be expanded into the eigenfunctions $u_{  k}(x)$, such that
\begin{align}
\mathcal{L}(x,v_{  x})=&\sum_{k=1}^\infty a_{  k} \sum_{l=1}^k \left[ \underline b_{  \, l, k}(v_{  x}) \right. \nonumber \\ &+ \left.\underline c_{  \, l, k}(v_{  x}) + \underline d_{  \, l}(v_{  x}) \, \delta_{  \, l, k} \right] \cos \sqrt{\lambda_{  k}} x
\end{align}
with an alternative set of velocity $v_{  x}$ dependent coefficients $\underline b_{  \, l, k}$, $\underline c_{  \, l, k}$, and $\underline d_{  \, l}$. These coefficients can also be written down explicitly, as shown below. For the kernel $\mathcal{R}$ in \eqref{operatorR} $\mathcal{R}$ we obtain straightforwardly 
\begin{align}
\mathcal{R}(x,v_{  x})=\sum_{k=1}^\infty \underline r_{  \, k}(v_{  x})\cos \sqrt{\lambda_{  k}} x.
\end{align}

Exploiting the orthogonality of the eigenfunctions, we ultimately find a linear system of equations for the unknown coefficients $a_{  k}$
\begin{align}
\frac{\lambda_{k} a_{  k}}{i\omega \mu_0 e}
+ 
\sum_{l=1}^k\left[
a_{  k} 
\int_{-\infty}^\infty \underline b_{  \, l, k}(v_{  x}) \, d v_{  x} + \underline D_{  \, l} \, \delta_{  \, l, k} \right]
\nonumber\\
= - \int_{-\infty}^\infty \underline r_{  \, k}(v_{  x}) d v_{  x}.
\end{align}
The known coefficients are cumbersome, but are given explicitly by
\begin{widetext}
\begin{align}
\underline b_{  \, l, k}(v_{  x})&=
	-\frac{e\, n \,v_{  x}^3\,\sqrt{\lambda_{  l} \lambda_{  k}} \,e^{-v_{  x}^2/(2 v_{  th}^2)}  
	\coth \left(\alpha L/v_{  x}\right)  \sin \sqrt{\lambda_{  l}} L \sin \sqrt{\lambda_{  k}} L }{
	\sqrt{\pi/2}\,L \,m \, v_{  th}\left(\alpha^2+v_{  x}^2 \lambda_{  l}\right)
	\left(\alpha^2+v_{  x}^2 \lambda_{  k}\right)},\\
\underline D_{  \, l}&=
	\int_{-\infty}^\infty \underline d_{  \, l}(v_{  x}) \, d v_{  x}
	=-\frac{ e \, n \, e^{\alpha^2/(2 \lambda_{  l} v_{  th}^2)} \,
	\text{erfc}\left[\alpha / (\sqrt{2 \lambda_{  l}} v_{  th
	})\right]}{\sqrt{2\lambda_{  l} /\pi} \,m \, v_{  th} },\\
\underline r_{  \, k}(v_{  x})&=
	-\frac{ i \omega B_0 \,e\, n \,e^{-v_{  x}^2/ (2 v_{  th}^2)} 
	\left[\left(1+e^{\alpha L /v_{  x}}\right) \alpha^3+\left(e^{\alpha L /v_{  x}}-1\right) 
	\left(v_{  x} \sqrt{ \lambda_{  k}} \right)^3 
	\sin \sqrt{ \lambda_{  k}} L \right]}{
	\sqrt{\pi/2}\, L \, m \, v_{  th}\,\lambda_{  k}\, \alpha^2   \left(1+e^{ \alpha L /v_{  x}}\right)  
    \left(\alpha^2+v_{  x}^2 \lambda_{  k}\right)}.
\end{align}
\end{widetext}
Here, the fact that the symmetric integration of any arbitrary function $g(v_{  x})$ over $v_{  x} \in (-\infty, \infty)$ is equal to the integral over its even contribution only has been used. In consequence, the integrals $\underline C_{  \, l, k}$ vanish after integration of $\underline c_{  \, l, k}(v_{  x})$ over $v_{  x}$, because $\underline c_{  \, l, k}(v_{  x})$ is an odd function in $v_{  x}$. In addition, the coefficients $\underline d_{  \, l}(v_{  x})$ have already been integrated analytically in $v_{  x}$, thus represented by $\underline D_{  \, l}$.

After numerical integration of the coefficients $\underline b_{  \, l, k}(v_{  x})$ and $\underline r_{  \, k}(v_{  x})$ over $v_{  x}$ and solving for $a_{  k}$, we obtain $E_{  h}$, which is the solution to the electric field equation  \eqref{rwp} subject to \eqref{rwbc1}--\eqref{rwbc2}. The total electrical field $\tilde E$ can then be calculated from \eqref{ansa} and consequently the magnetic field $\tilde B$ and the current density $\tilde j$. We have 
\begin{align}
\tilde E&=i\omega (L-x)B_0 + \sum_{k=1}^\infty a_{  k} \cos \sqrt{\lambda_{  k}} x ,\\
\tilde B&=-\frac{1}{i\omega}\frac{\partial \tilde E}{\partial x},\\
\tilde{\! j}&= \frac{1}{i\omega\mu_0}\frac{\partial^2 \tilde E}{\partial x^2}.
\end{align}


\section{Results and discussion}

\subsection{Analytical results}

In this subsection, the analytic solution to the one-dimensional plasma slab domain depicted in figure~\ref{fig01} is self-consistently evaluated for generic parameters of an ICP discharge. The length of the plasma is set to $L=6$ cm. The plasma density and the electron temperature is chosen to be $n=5\times 10^{11}$  cm$^{-3}$ and 4 eV, respectively. We chose a simple argon plasma model, where the electron-neutral momentum transfer collision frequency is specified as $\nu_{  m}=K_{  m} n_{  g}$, with the rate coefficient $K_{  m}=10^{-13}$~m$^3$s$^{-1}$ and the neutral gas density as a function of gas pressure $n_{  g}$(m$^{-3}$)=$3.3\times10^{22}\,p$(Torr). The magnetic field at the interface between the vacuum and the plasma at $x=0$ is set to $B_0=10^{-6}$ Vsm$^{-2}$. The angular driving frequency $\omega$ and the pressure $p$(Torr) are parameters which are varied, and thus specified in the respective paragraphs.

\begin{figure}[t]
\includegraphics[width=8cm]{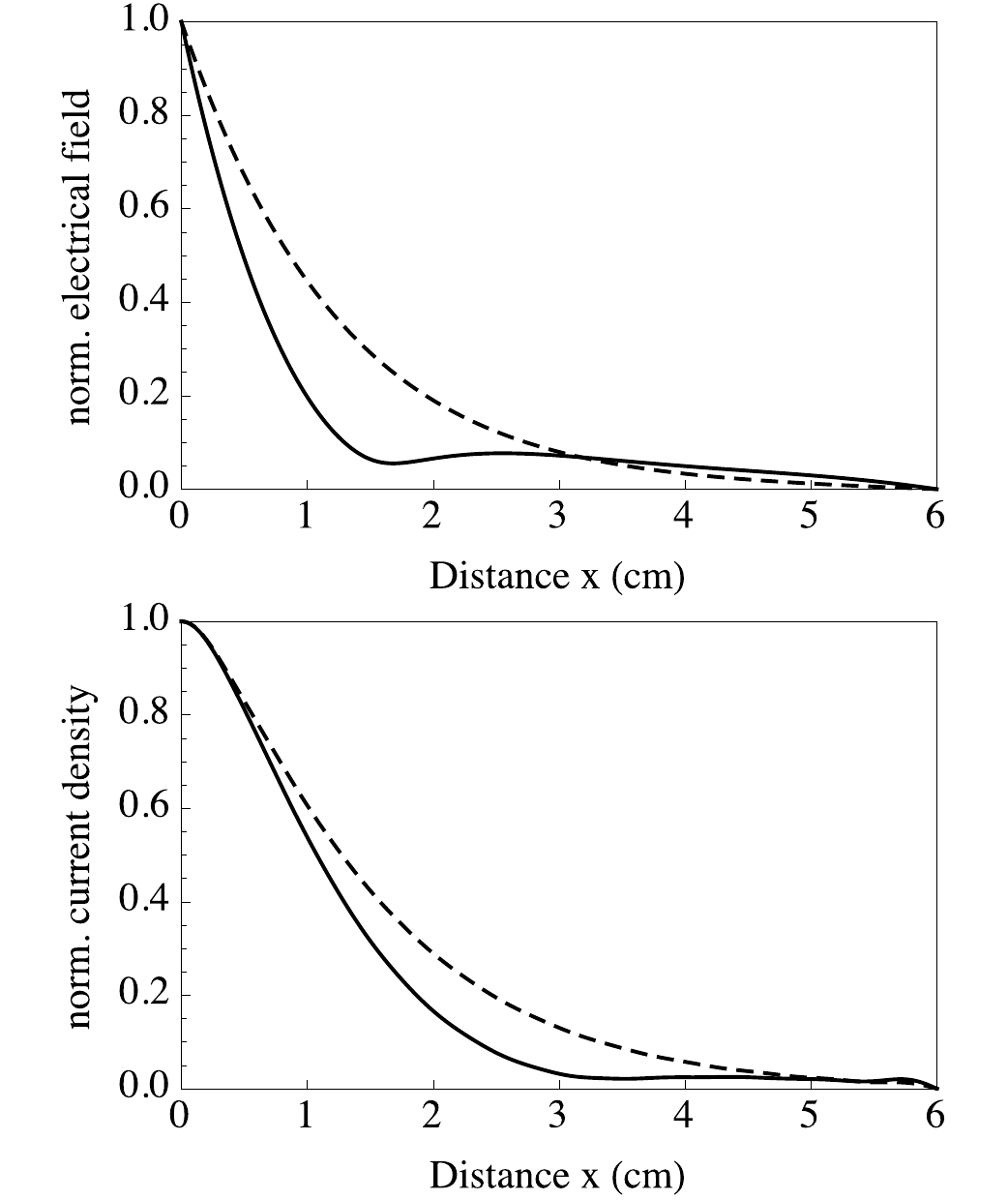}
\caption{Electrical field (above) and current density (below) for $\omega=2\pi\times 13.56$ MHz and different gas pressures: $p=$ 1 mTorr (solid lines), $p=$ 45 mTorr (dashed lines).}
\label{fig02}
\end{figure}

Figure~\ref{fig02} shows the magnitude of both, the electrical field (above) and the current density (below) for $\omega=2\pi\times 13.56$ MHz and two different gas pressure. In the high pressure regime $p=$ 45 mTorr (i.e., the local regime) indicated by dashed lines an exponential decrease of both fields is observed, which is in fact typical for the classical skin effect. In the non-local regime at $p=$ 1 mTorr, the electrical field as well as the current density reveal a non-monotonic decrease, which is typical for the anomalous skin effect. It is characterized by the appearance of local minima and maxima in field's magnitudes.  

\begin{figure*}[t]
\includegraphics[width=16cm]{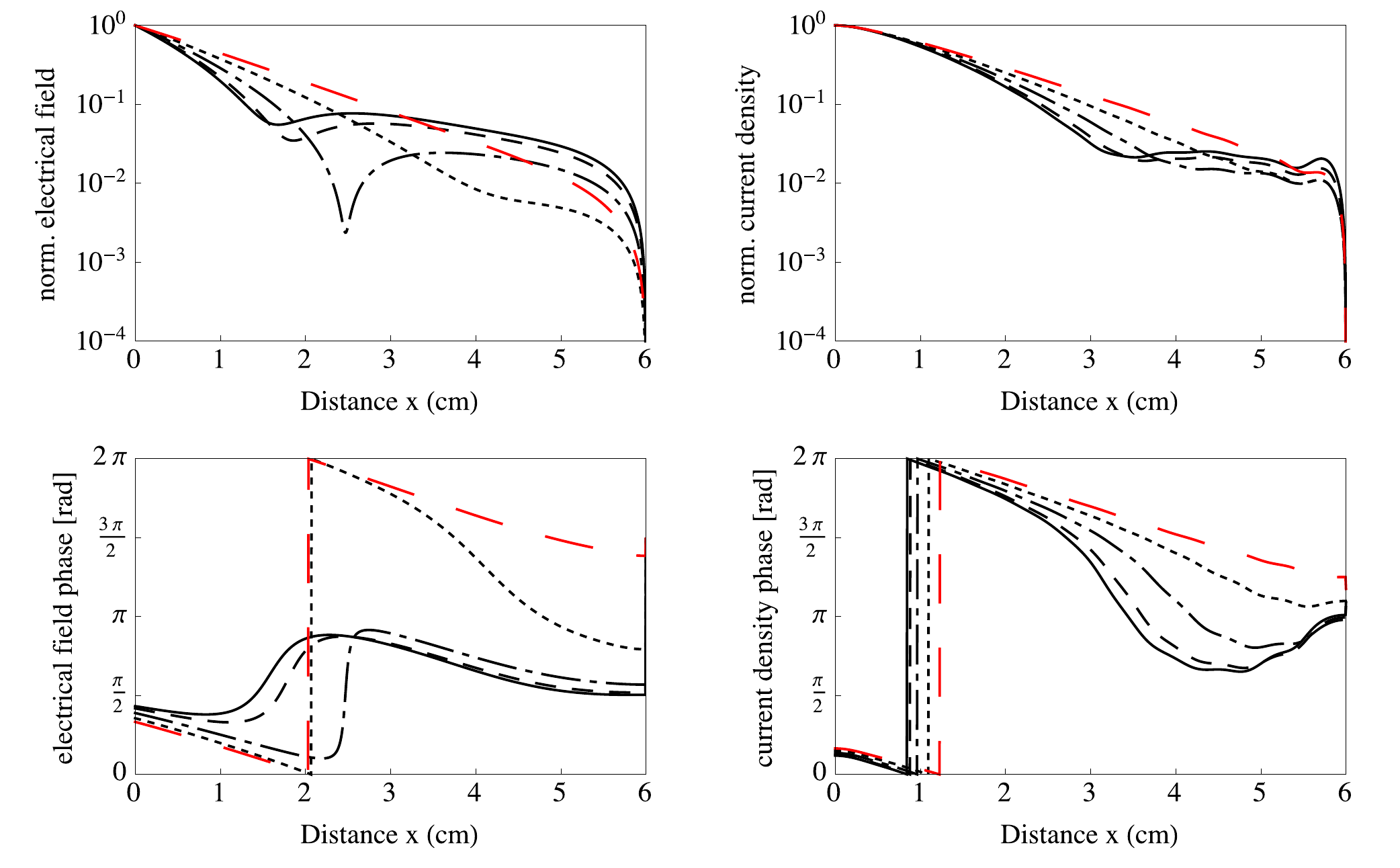}
\caption{Magnitude (top) and phase (below) of the electrical field (left) and the current density (right) for $\omega=2\pi\times 13.56$ MHz and different gas pressures: $p=$ 1 mTorr (solid lines), $p=$ 5 mTorr (dashed lines), $p=$ 15 mTorr (dash-dotted lines), and $p=$ 30 mTorr (dotted lines), and $p=$ 45 mTorr (red, long-dashed line).}
\label{fig03}
\end{figure*}

The appearance of the non-monotonic spatial distribution of the fields, moreover, translates into the complex phase of the time-harmonic observables. This can be seen in figure~\ref{fig03}, where the absolute values (above) as well as the phases (below) of both the electric field (left) and the current density (right) are shown for varying gas pressures. For all pressures the more moderate decay of the current density distribution compared to the electric field decay is obvious. This demonstrates the effect of the current diffusion due to thermal electron motion (i.e., a warm plasma effect). The individual phases are themselves directly linked to their penetration depth. The profiles of the phases are correspondingly different for the electric field and current density. The higher the pressure (i.e., more local) the more similar they are. The fact that they are different suggests distinct mechanisms of propagation of the electric field and the current density.

\begin{figure}[t]
\includegraphics[width=8cm]{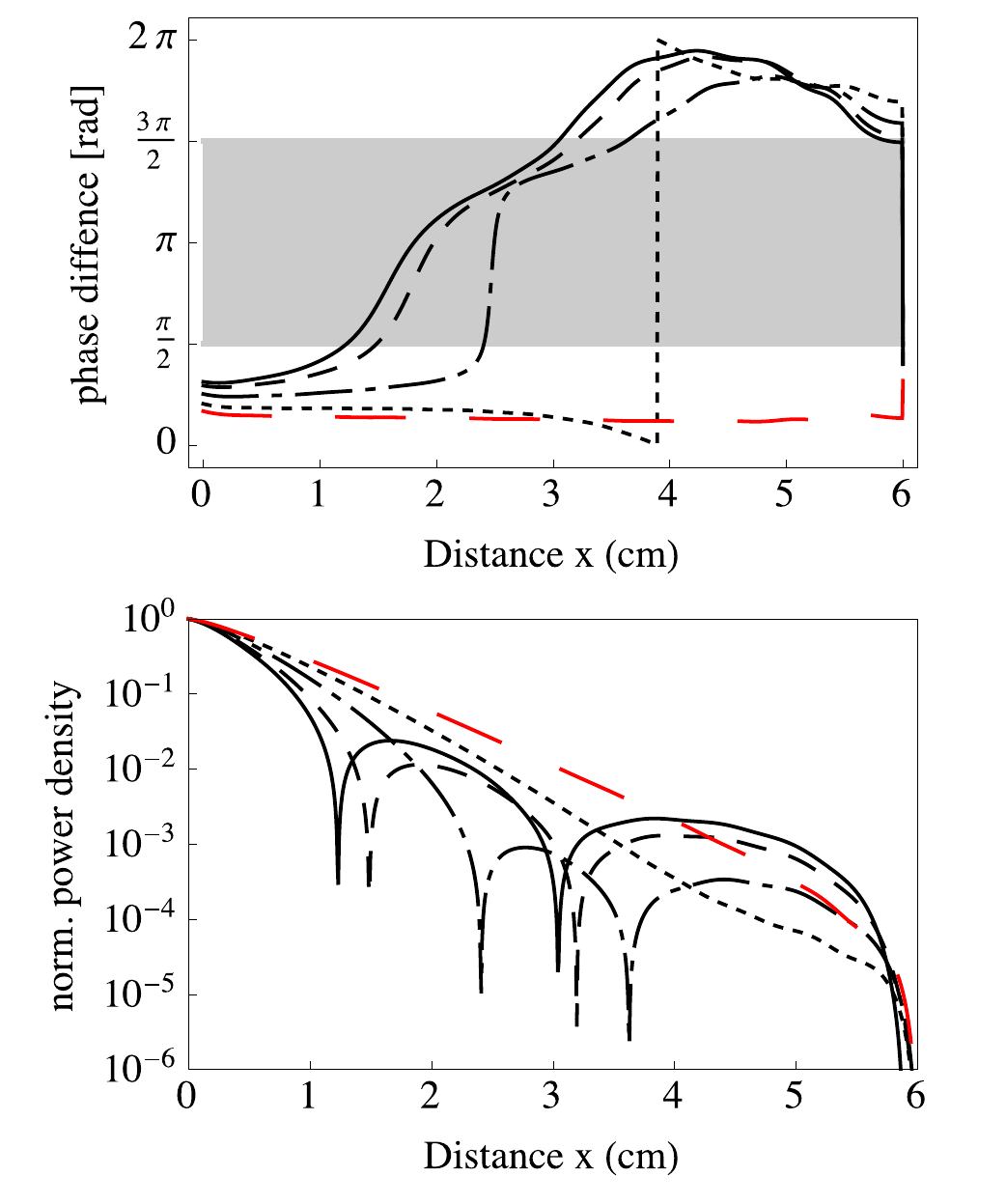}
\caption{Phase difference (top) and power density (below) for $\omega=2\pi\times 13.56$ MHz and different gas pressures: $p=$ 1 mTorr (solid lines), $p=$ 5 mTorr (dashed lines), $p=$ 15 mTorr (dash-dotted lines), and $p=$ 30 mTorr (dotted lines), and $p=$ 45 mTorr (red, long-dashed line). Highlighted in gray is the interval of negative power deposition.}
\label{fig04}
\end{figure}

It should be recalled that the spatial distributions of the electric field and the current density are comparable for the normal skin effect -- scaled in magnitude, but identical in phase. This is indicated by figure~\ref{fig03} (red, long-dashed line). Both phases are simply shifted and their phase difference $\Delta \phi$ remains less than $\pm \pi/2$. It is zero in the limit of purely Ohmic conduction and close to zero in the local regime (cf., $p=$ 45 mTorr).

\begin{figure*}
\includegraphics[width=16cm]{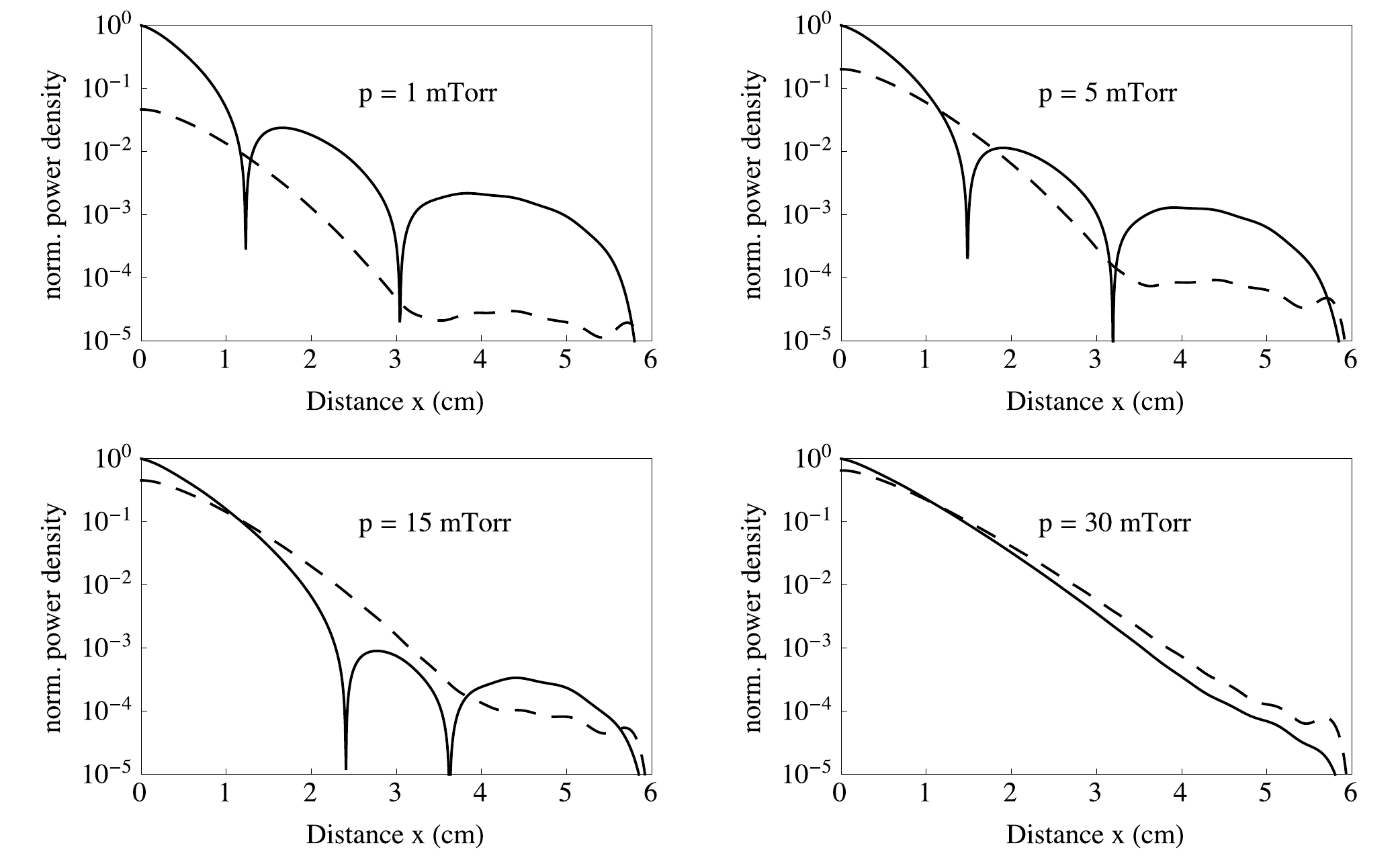}
\caption{Total (solid lines) and Ohmic (dashed lines) power density for $\omega=2\pi\times 13.56$ MHz and different gas pressures: $p=$ 1 to 30 mTorr.}
\label{fig05}
\end{figure*}

\begin{figure*}
\includegraphics[width=16cm]{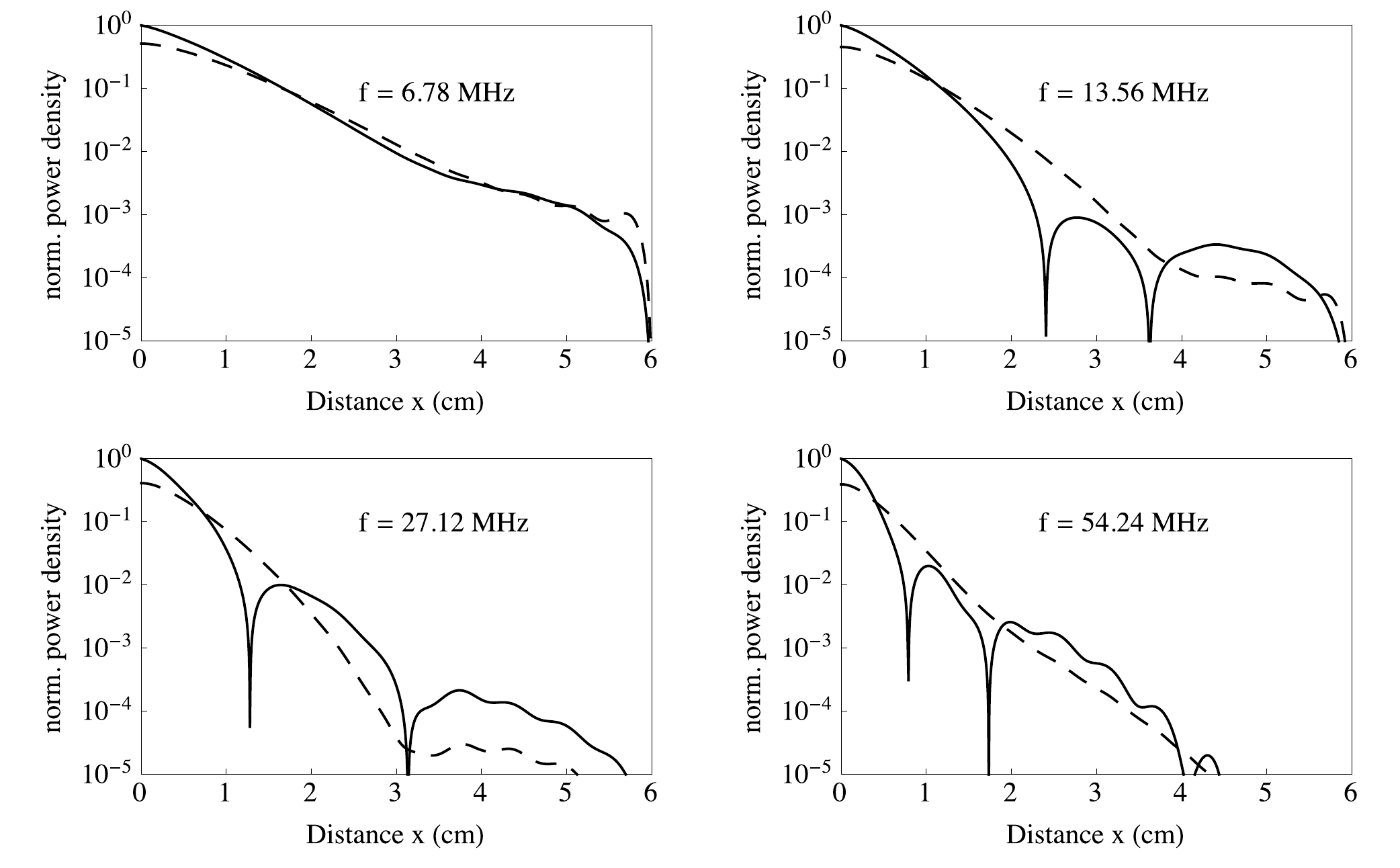}
\caption{Total (solid lines) and Ohmic (dashed lines) power density for $p=$ 15 mTorr and different rf frequencies $\omega=2\pi\times 6.78$ to $2\pi\times 54.24$ MHz.}
\label{fig06}
\end{figure*}

The total power density $p_{  abs}$ deposited can be generally calculated from
\begin{align}
\label{total}
p_{  abs}=\frac{1}{2} {  Re}\ \tilde{\! j} \tilde E^*.
\end{align}
In the local regime only positive (Ohmic) power deposition is to be recognized. In contrast, in the non-local regime phase differences $\Delta \phi$ between electric field and current density with values in the range $\pi/2 < \Delta \phi < 3\pi/2$ are observed. This is depicted in figure~\ref{fig04} (above). Here the above mentioned range is indicated by the gray region. A phase shift within these limits represents negative power deposition (figure~\ref{fig04}, below). Whenever the phase difference crosses the specified interval, the power deposition changes its sign. In the logarithmic plot given below, the zero-crossing of the power density diverges. The negative power deposition is clear from the mathematical point of view. The distribution of power density deposited throughout its propagation into the plasma is calculated by $p_{  abs}= 1/ 2 \,\, |\tilde{\! j}| \, |\tilde E| \, \cos \Delta\phi$. In terms of a polar coordinate representation within the complex plane the real part is clearly negative for the above mentioned phase difference interval. Although clear from a mathematical standpoint, negative power deposition is a quite remarkable physical phenomenon nevertheless.

Ohmic power deposition due to collisions of electrons with the neutral gas background is always positive. Given a conductivity $\sigma$, the purely Ohmic power density contribution can be approximated. In the limit of a cold collisional plasma (assuming locality), the plasma conductivity is $\sigma=e^2 n_e/m_e(\nu_{  m} + i\omega)$. Correspondingly, from the local version of Ohm's law the power density can in principle be evaluated from either the current density or the electric field. However, given that Ampere's law ties the current density to the curl of the magnetic field irrespective of any material model (i.e., Ohm's law), it seems more reasonable to approximate the Ohmic power density from the non-local current density (in favor of the non-local electric field). It can thus be calculated from
\begin{align}
\label{ohmic}
p_{  ohm}= \frac{1}{2} {  Re} (\sigma^{-1}) \ \tilde{\! j}\ \tilde{\! j}^*.
\end{align}

The power density distribution deposited is given in figure~\ref{fig05} for various gas pressures at a constant frequency of the electromagnetic field of $f = 13.56$ MHz. The total deposited power density is represented by solid lines, while the Ohmic contribution is indicated by dashed lines. Clearly, the higher the pressure, the better the local power density calculated from expression \eqref{ohmic} approximates the exact non-local power density equation \eqref{total} deposited. For the case of $p=$ 30 mTorr both are nearly the same. Only in the low pressure cases $p \leq$ 15 mTorr the remarkable feature of negative power deposition is apparent. In these situations, power is removed from certain regions of negative power deposition and transferred to regions of positive power deposition. This implies that a mechanism acts to turn the positive sign of net power deposition negative. This is exactly due to collisionless heating. Electrons gain energy throughout certain periods of time, but experience a randomizing collision only much later after having traversed to a different position within the discharge. The collective effect is clearly non-local in position space. For the most obvious case of $p=$ 1 mTorr, negative power deposition is observed in the interval $1.2 \lesssim x \lesssim 3$ cm. Of course this phenomenon is only visible for the total power density $p_{  abs}$, the Ohmic power density is $p_{  ohmic} \geq 0$ in the whole discharge.

For all cases it can be seen that the energy is predominantly absorbed near the plasma boundary and that deeper within the plasma alternating regions of negative and positive power deposition appear. The penetration depth as well as the phase structure are non-trivially linked to the degree of non-locality. A phenomenon previously characterized (among others) by \cite{turner1993} who discusses the peculiar parameter space in which collisionless heating efficiently takes place. In particular: (i) That the inhomogeneity within the discharge be large enough allowing the electrons to experience regions of different field amplitude. And more importantly, (ii) it is essential that within one rf period the electron motion are sufficiently fast, such that they traverse a distance larger then the depth of field penetration. This is intrinsically linked to the mean electron energy (e.g., the electron temperature $T_{  e}$) and the collision frequency $\nu_{  m}$.

Because the time for a ``warm'' electron to proceed without experiencing a significant change of the electromagnetic field is largely dependent of the rf frequency, a similar phenomenon as for different pressures can be observed for various frequencies $\omega$ of the penetrating electromagnetic field while keeping the pressure fixed at $p=$ 15 mTorr. This is depicted in figure~\ref{fig06}. Also in this case, the occasions and extent of regions with negative power deposition strongly vary, dependent on the actual frequency and corresponding phase relation of the penetrating electric field and current density. As apparent from the top left subfigure in figure~\ref{fig06}, for the case of $\omega=2\pi\times 6.78$ MHz no negative power deposition is observed. The net power absorption is very close to the Ohmic power density absorbed in local approximation. Only for larger frequencies of $\omega \gtrsim 2\pi\times 13.56$ MHz negative power deposition is present. With increasing frequency the regions of negative power deposition are quenched closer to the driven boundary. These results are in good agreement with previous model results and experimental data.\cite{kolobov1997,godyak1997,godyak1998,tyshetskiy2002}

\subsection{Comparison with Particle in Cell}

For comparison, it is instructive to review the identical discharge setup also in terms of an kinetic simulation approach. For this purpose the particle in cell/Monte Carlo collisions (PIC/MCC) approach is utilized, taking into account the full-wave transversal electromagnetic fields (i.e., $E_{  y}$ and $B_{  z}$). An extended version of the \emph{yapic} code, which has been previously benchmarked against other PIC implementations, is utilized.\cite{turner2013} Within the simulation the configuration space is reduced to one dimension, but all three components in velocity space are retained (i.e., 1D3V). The essentials of the proposed simulation model are presented elsewhere.\cite{trieschmann2013,turner2013} The employed modifications can be summarized as follows: (i) While the transient electron dynamics are maintained, for the ions a homogeneous and constant density is imposed within the plasma slab. (ii) With the same reasoning, imposing the assumption of a quasi-neutral plasma bulk, the longitudinal electric field $E_{  x}$ is virtually zero using the expression for the ambipolar field $\vec{E} = -\frac{T_e}{n_i e} \nabla n_i \equiv 0$ for $n_i = {  const}$.\cite{joyce1997,sydorenko2005} (Principally, the \emph{longitudinal} electric field could have been also obtained from Poisson equation based on the electrostatic approximation.) Both assumptions assure that the plasma conditions are as similar as possible to the ones assumed within the analytic model. A fortuitous side effect is a greatly diminished statistical noise level that is inherent to PIC. (iii) The full-wave \emph{transversal} electromagnetic fields $E_y$ and $B_z$ are incorporated by means of a standard finite difference time domain (FDTD) approach.\cite{yee1966,taflove2005}

\begin{figure}
\includegraphics[width=8cm]{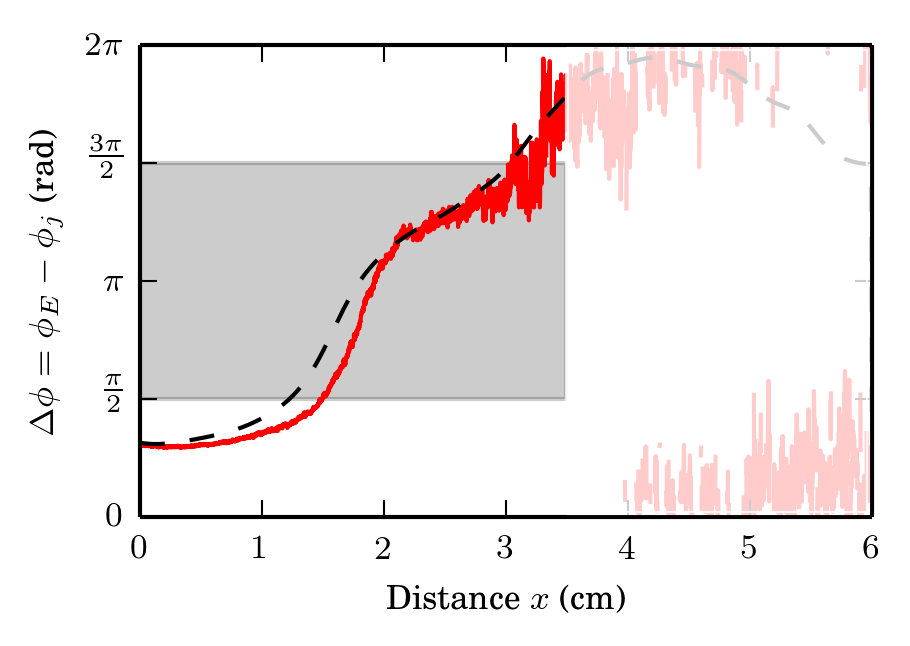}
\caption{Phase difference for the analytical model (dashed black line) and obtained from PIC (solid red line) for $\omega=2\pi\times 13.56$ MHz and a gas pressures of $p=$ 1 mTorr. Highlighted in gray is the interval of negative power deposition. The region for $x > 3$ cm is not reliable due to too strong statistical noise.}
\label{fig07}
\end{figure}

Next, PIC simulation results for the mentioned discharge scenario with a frequency $f = 13.56$ MHz and at a pressure of $p = 1$ mTorr shall be discussed. Figure~\ref{fig07} presents the correspondingly obtained spatial profile of the relative phase (solid red) between the transversal electric field and the current density (i.e., the parallel contribution) in comparison with the result from the analytical model (dashed black). Again shaded in gray is the interval $\pi/2 < \Delta \phi < 3\pi/2$ of negative power deposition. As stands out, a remarkable similarity in the results is observed, despite the very different level of complexity of the compared kinetic models. The spatial profile is qualitatively nearly equivalent. The observed physical consequences due to the anomalous skin effect is clearly inherent to both models. Interestingly, there are slight deviations in numbers, in particular as regards the position of phase transition from positive to negative power deposition at $x \approx 1.3$ cm. It is, however, important to acknowledge that the phase and therewith the exact position of the transition is a rather sensitive measure. This will be addressed shortly.

The phase transition is intrinsically connected to the power deposition as depicted in figure~\ref{fig08} for both PIC (solid red) as well as the analytical model (absolute power density: dashed black; Ohmic power density: dash-dotted blue). It is apparent that the position of phase transition toward negative power deposition has a one to one correspondence with the phase difference between the electric field and the current density. The transition to a negative real part in polar representation is reflected by a zero crossing of the absolute deposited power (i.e., a singularity in the semi-log plot shown). As expected from the phase difference, the singularity obtained from PIC is slightly shifted with respect to the analytical model results. Besides this shift, the principle physical dynamics as well as the quantitative dependence are resembled closely. In the frame of a comparison, two aspects seem quite peculiar:

\begin{figure}
\includegraphics[width=8cm]{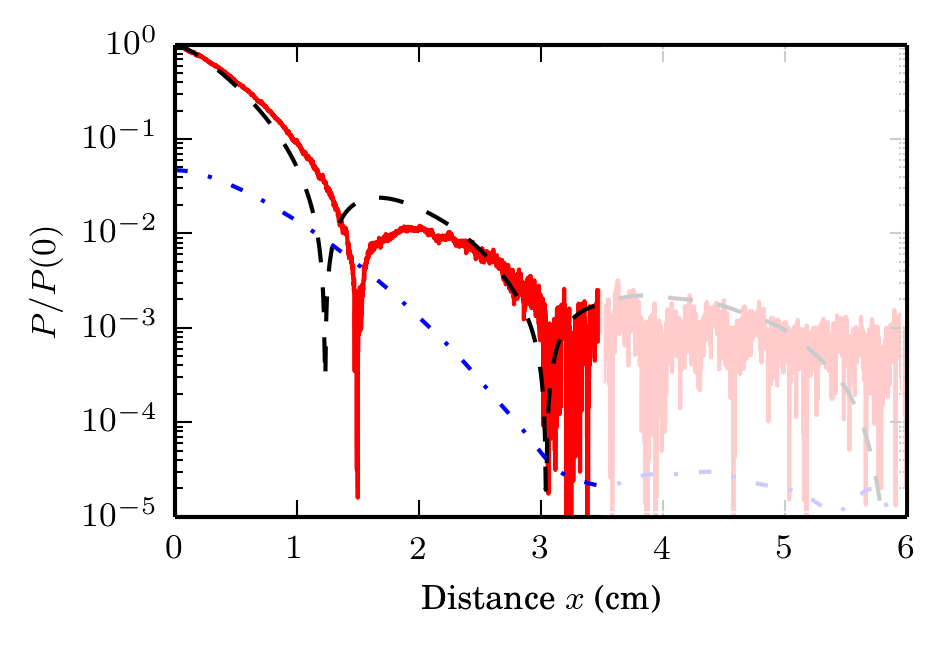}
\caption{Absolute (dashed black line) and Ohmic (dash-dotted blue line) power density for the analytical model and obtained from PIC (solid red line) for $\omega=2\pi\times 13.56$ MHz and a gas pressures of $p=$ 1 mTorr. Highlighted in gray is the interval of negative power deposition. The region for $x > 3$ cm is not reliable due to too strong statistical noise.}
\label{fig08}
\end{figure}

Firstly, the level of statistical noise observed in the PIC results. Being inherent to all Monte Carlo models, accounting for the essential stability and accuracy considerations, these are of rather cosmetic concern. In terms of an evaluation and interpretation, caution is suggested when the signal is too noisy as for instance in figure~\ref{fig08}, for distances larger than $x \approx 3$ cm. The electric field strength is maximum at the left interface and strongly damped while penetrating the plasma. This is due to the low excitation frequency in comparison with the electron plasma frequency. Consequently, the signal greatly diminishes and thus the signal to noise ratio substantially increases from the plasma interface into the plasma (i.e., left to right).

Secondly, deviations due to contrasting model assumptions. While the goal was to differentiate the two models as least as possible, a few differences are intrinsic to the respective model. In approximate order of importance: (i) Within the analytic model collisions are irrespective of energy taken into account by a constant collision frequency. In comparison to the collisional processes and energy dependent cross sections utilized in the PIC model, this is a rather crude approximation.\cite{phelps1994,lxcat2012} (ii) In the analytic model the force term within the Boltzmann equation is reduced to the electric field force. In contrast, in the PIC model the complete Lorentz force term $\vec F = q ( \vec E + \vec v \times \vec B )$ is retained. (iii) The analytical model assumes the displacement current $\varepsilon_0 \frac{\partial \vec E}{\partial t}$ to be negligible. It is, however, included in the full-wave PIC model. As the total current density within the plasma (the sum of the conduction and the displacement current density) is vastly dominated by the conduction current, this assumption seems quite unimportant. (iv) The analytic model is developed around a linear perturbation. Nonlinear effects are consequently excluded. Within PIC nonlinear effects are intrinsically included. However, for the investigations performed the field strength has been intentionally chosen small enough, so that nonlinear effects do not play any role.\cite{froese2009}

To conclude with the comparison of the analytic and the PIC model, both appear to be in remarkable agreement given the detailed differences between the two model approaches. It is to notice, however, that a great difference between the two models lies within the time and computational effort required for the solution. While the analytical model is evaluated within seconds, typical runs for the (non optimized) PIC code took about two months to only provide a \emph{fair} statistical basis. The analytical model therefore facilitates exceptional accuracy of the calculations results paired with a tremendously reduced computational burden.

\section{Conclusions}

The negative power absorption in low pressure radio frequency plasmas is investigated by means of an analytical model which couples Boltzmann's equation and the quasi-stationary Maxwell's equation. The formulated boundary value problem is of Sturm-Liouville type and is solved exploiting standard Hilbert space methods. An explicit solution for both, the electric field and the distribution function of the electrons has been found for the first time for a bounded unsymmetrical discharge configuration. Particularly, the anomalous skin effect with its peculiarities, e.g., the non-monotonic decay of the field distribution and the effect of phase mixing is discussed. The analytical solution is compared with results from particle in cell simulations. Here a one-dimensional electromagnetic full wave model has been developed and implemented. A comparison of the analytical and the numerical results show an excellent agreement.


\section*{Acknowledgments}
The authors gratefully acknowledge the support by the Deutsche
Forschungsgemeinschaft in the frame of Collaborative Research Centre TRR 87. Valuable discussions with and support from Prof. T. Eisenbarth, Dr. M. Lapke, and Prof. R.P. Brinkmann is gratefully acknowledged.

\end{document}